\documentclass[aps, prl, reprint, showpacs, twocolumn, amsfonts, amsmath, amssymb, superscriptaddress, floatfix]{revtex4-1}
\usepackage[OT1]{fontenc}
\usepackage[latin1]{inputenc}

\usepackage{color}
\usepackage{graphicx}
\usepackage{ulem}
\usepackage[colorlinks,bookmarks=false,citecolor=blue,linkcolor=red,urlcolor=black]{hyperref}
\newcommand{\kbar}{\mathchar'26\mkern-9mu k}

\usepackage[english]{babel}
\begin{document}

\preprint{This line only printed with preprint option}

\title{Anderson transition of Bogoliubov quasiparticles in the quasiperiodic kicked rotor}

\author{Beno\^{i}t Vermersch}
\altaffiliation[Present Address: ]{Institute for Quantum Optics and Quantum Information of the Austrian Academy of Sciences, A-6020 Innsbruck, Austria,
benoit.vermersch@uibk.ac.at}

\affiliation{Laboratoire de Physique des Lasers, Atomes et Mol{\'e}cules, Universit{\'e}
Lille 1 Sciences et Technologies, CNRS; F-59655 Villeneuve d'Ascq
Cedex, France}

\author{Dominique Delande}

\affiliation{Laboratoire Kastler-Brossel, UPMC-Paris 6, ENS, CNRS; 4 Place Jussieu,
F-75005 Paris, France}

\author{Jean~Claude Garreau}

\affiliation{Laboratoire de Physique des Lasers, Atomes et Mol{\'e}cules, Universit{\'e}
Lille 1 Sciences et Technologies, CNRS; F-59655 Villeneuve d'Ascq
Cedex, France}

\homepage{http://www.phlam.univ-lille1.fr/atfr/cq}

\begin{abstract}
We study the dynamics of Bogoliubov excitations of a Bose-Einstein
condensate in the quasiperiodic kicked rotor. In the weakly interacting
regime, the condensate is stable and both the condensate and the excitations undergo a phase transition from a quasilocalized 
to a diffusive regime. The corresponding critical exponents are identical for 
the condensate and the excitations, and
compare very well with the value $\nu\approx1.6$ for non-interacting particles.
\end{abstract}

\maketitle

Ultracold atoms are clean, controllable, and flexible 
systems whose dynamics can be modeled from first principles.
Interacting ultracold bosons are often well described in
the frame of a mean-field approximation leading to the Gross-Pitaevskii
equation (GPE)~\cite{Stringari:BECRevTh:RMP99}, which is useful
in a wealth of situations of experimental interest: Superfluidity
and vortex formation~\cite{Madison:VortexFormationBEC:PRL00}, chaotic
behavior~\cite{Thommen:ChaosBEC:PRL03,*Fallani:InstabilityBEC:PRL04,*Gattobigio:ChaoticPotentialsMatterWave:PRL11},
soliton propagation~\cite{Khaykovich:MatterWaveBrightSoliton:S13},
etc. Ultracold atom systems are thus increasingly used to realize
simple models that are inaccessible experimentally
in other areas of physics~\cite{Bloch:QuantumSimulationsUltracoldGases:NP14}.

Ultracold gases in a disordered 
optical potential have been used as an emulator for the Anderson model~\cite{Anderson:LocAnderson:PR58},
allowing the direct observation of the Anderson 
localization~\cite{Kondov:ThreeDimensionalAnderson:S11,*Jendrzejewski:AndersonLoc3D:NP12,*Semeghini:MobilityEdge3DAnderson:arXiv14}. 
The quantum kicked rotor (QKR), obtained by placing cold
atoms in a pulsed standing wave, is also a (less obvious) emulator
for the Anderson physics~\cite{Casati:LocDynFirst:LNP79,*Fishman:LocDynAnders:PRL82}: It displays
\emph{dynamical localization}, a suppression of chaotic diffusion in the 
momentum space, recognized to be equivalent to the Anderson localization. Recent
studies suggest that interactions (treated in the frame of the GPE)
lead to a progressive destruction of the dynamical localization, which
is replaced by a subdiffusive regime~\cite{Shepelyansky:KRNonlinear:PRL93,*Gligoric:InteractionsDynLocQKR:EPL11,*Rebuzzini:EffectsOfAtomicInteractionsQKR:PRA07,
Cherroret:AndersonNonlinearInteractions:PRL14}
in analogy with what is numerically observed for the 1D Anderson model
itself~\cite{Pikovsky:DestructionAndersonLocNonlin:PRL08,
*Laptyeva:DisorderNonlineChaos:EPL10,*Vermersch:AndersonInt:PRE12}.

Applying standing-wave pulses (kicks) to a Bose-Einstein condensate
(BEC) may lead to a dynamical instability which transfers atoms from
the condensed to the non-condensed fraction,
a phenomenon which is not described by the GPE. The most common ``higher
order'' approximation in this context is the Bogoliubov-de Gennes
(BdG) approach~\cite{Bogoliubov:TheTheoryOfSupefluydity:JPURSS47,*deGennes:SuperconductivityOfMetals:66}. 
The BdG theory considers ``excitations'' -- described as independent bosonic
quasiparticles --  of the Bose gas, and thus indicates how (and how
much) it differs from a perfectly condensed gas.
It has been applied both to describe the dynamical
instability of the periodic kicked rotor~\cite{Zhang:TransitionToInstabilityKickedBEC:PRL04,*Billam:CoherenceAndInstabilityInADrivenBEC:NJP12,*Reslen:DynamicalInstabilityInKickedBEC:PRA08}
and to study a one-dimensional weakly interacting BEC~\cite{Lugan:AndersonLocalizationBogolyubov:PRL07,*Gaul:BogoliubovExcitationsOfDisorderedBEC:PRA11}
in a disordered potential; it was found in the latter case that the quasiparticles
may also display Anderson localization. Interestingly, a modified version of
the QKR, the \emph{quasiperiodic} kicked rotor (QPKR), in the absence of interactions, emulates the dynamics of
a \emph{3D} Anderson-like model, and displays the Anderson metal-insulator
transition~\cite{Casati:IncommFreqsQKR:PRL89}. With this system  a rather complete theoretical and experimental study of this transition
has been performed
~\cite{Chabe:Anderson:PRL08,Lemarie:CriticalStateAndersonTransition:PRL10,*Lopez:PhaseDiagramAndersonQKR:NJP13,
Lopez:ExperimentalTestOfUniversality:PRL12}.
In the present work we apply the BdG approach to the QPKR,
both to study the stability of the condensate and the dynamics of its Bogoliubov excitations. 
We show that for weak enough interactions, the condensate remains stable for experimentally relevant times, and 
that the Bogoliubov quasiparticles also display the Anderson phase transition.

A kicked rotor is realized by submitting ultracold atoms
to short kicks of a standing wave at times separated by a constant
interval $T$. If such kicks have a constant amplitude, one realizes
the standard (periodic) kicked rotor; which exhibits dynamical localization
~\cite{Casati:LocDynFirst:LNP79,Moore:AtomOpticsRealizationQKR:PRL95}.
If the amplitude of the kicks is modulated with a quasiperiodic function
$f(t)=1+\epsilon\cos\left(\omega_{2}t+\varphi_{2}\right)\cos\left(\omega_{3}t+\varphi_{3}\right)$,
where $\omega_2 T$, $\omega_3 T$  and $\kbar\equiv 4\hbar k_L^2T/M$ (the reduced Planck constant) are incommensurable ($k_L$ is the 
wave-vector of the 
standing wave and $M$ is the mass of the atoms), the QPKR is obtained~\cite{Casati:IncommFreqsQKR:PRL89}. In the absence of particle-particle
interactions, the Hamiltonian of the QPKR, in conventional normalized
units~\cite{Moore:AtomOpticsRealizationQKR:PRL95,Lemarie:AndersonLong:PRA09}, is:
\begin{equation}
H=\frac{p^{2}}{2}+K\cos x\ f(t)\sum_{n\in\mathbb{N}}\delta(t-n).\label{eq:Hqpkr}
\end{equation}
where $K$ is proportional to the average standing wave intensity.
In such units the time interval between kicks is $T=1$,
lengths are measured in units of $(2k_L)^{-1}$.
Throughout this work
we take $\omega_{2}=2\pi\sqrt{5}$, $\omega_{3}=2\pi\sqrt{13}$ and
$\kbar=2.89$ corresponding to typical experimental values~\cite{Chabe:Anderson:PRL08,Lemarie:CriticalStateAndersonTransition:PRL10,Lopez:ExperimentalTestOfUniversality:PRL12}.
In the absence of interactions the QPKR displays, for low values of $K$ and $\epsilon$, 
dynamical localization at long times  (i.e. $\left\langle p^{2}\right\rangle \sim \mathrm{constant}$); 
for $K\gg1$, $\epsilon\approx1$ one observes a
diffusive regime $\left\langle p^{2}\right\rangle \sim t$, and in between
there is a critical region which displays a subdiffusive behavior
$\left\langle p^{2}\right\rangle \sim t^{2/3}$~\cite{Lemarie:AndersonLong:PRA09}. 

We use in the present work a model slightly different of the experimentally
realized QPKR; we consider a  sinusoidal potential ``folded"
over one spatial period of the standing wave; in such case $p$ becomes
a discrete variable $p=\kbar l$, with $l\in \mathbb{Z}$. Mathematically, this is
equivalent to use one-period ($2\pi$ in normalized units) spatial
periodic boundary conditions.
In the presence of weak interactions that are modeled by a mean-field nonlinear potential, the critical and the diffusive regime are not affected whereas the localized regime is replaced by a subdiffusive one $\langle p^2 \rangle \sim t^\alpha$, with $\alpha\sim0.4$ \cite{Cherroret:AndersonNonlinearInteractions:PRL14,Ermann:DestructionOfAndersonLocNonlinearity:JPAMG14}. In the following, we consider the weakly interacting regime and short time scales so that this effect is negligible; we shall thus use the term ``quasilocalized" to characterize this phase.

We emphasize that in our model there is no spatial dilution of the boson gas, and the average nonlinear potential, which is proportional to the density, does not vary with time. 
This is not the case in the usual experimental realization of the QPKR, where the atom cloud diffuses with time, causing a significant diminution of the spatial density; once the system is diluted,  the nonlinearity does not play anymore an important role. Our model is thus expected to catch more clearly the
physics in presence of the nonlinearity. One can use a torus-shaped confining optical potential in order to 
realize experimentally such a geometry~\cite{Ramanathan:SuperflowToroidalBEC:PRL11}.

In this quasi-1D geometry, we take interactions into account via
 the particle-number-conserving Bogoliubov formalism~\cite{Castin:LowTemperatureBECsInTimeDependent:PRA98}, at zero
temperature. The gas of interacting bosons is separated into two parts:
(i) The condensed part (or condensate) and (ii) the
non-condensed part (or ``excitations''). The condensate is governed by the Gross-Pitaevskii equation 
\begin{equation}
i\kbar\frac{\partial\phi(x)}{\partial t}=H\phi(x)+g|\phi(x)|^{2}\phi(x)\label{eq:evolphi}
\end{equation}
where the condensate wave function $\phi$ is normalized to unity: $\int_0^L |\phi(x)|^2 dx = 1$  ($L=2\pi$ is the system length) 
and the rescaled 1D interaction strength $g=2\kbar \omega_\perp a N$ is proportional to the $S$-wave scattering length $a$, the number of atoms $N$ and the transverse trapping frequency $\omega_\perp$. The non-condensed part is described in the Bogoliubov formalism as a set of 
independent bosonic quasiparticles, whose two-component state vector $(u_{k},v_{k})$, 
satisfying the normalization condition $\int_0^L \left[|u_k|^2(x) - |v_k|^2(x)\right]\  dx=1$,
evolves according to the equation:
\begin{equation}
i\kbar\partial_{t}\left[\begin{array}{c}
u_{k}\\
v_{k}
\end{array}\right]=\mathcal{L}\left[\begin{array}{c}
u_{k}\\
v_{k}
\end{array}\right].\label{eq:evol_uk}
\end{equation}

The operator $\mathcal{L}$ is a $2\times2$ matrix: 
\begin{eqnarray}
\mathcal{L} & = & \left[\begin{array}{cc}
Q\\
 & Q^\dagger
\end{array}\right]\mathcal{L_{\mathrm{GP}}}\left[\begin{array}{cc}
Q\\
 & Q^\dagger
\end{array}\right]\nonumber\\
\mathcal{L_{\mathrm{GP}}} & = & \left[\begin{array}{cc}
H+2g|\phi|^{2}-\mu(t) & g\phi^{2}\\
-g\phi*^{2} & -H-2g|\phi|^{2}+\mu(t)
\end{array}\right],
\label{eq:def_LGP}
\end{eqnarray}
where $\mathcal{L}_{\mathrm{GP}}$ is the usual Bogoliubov operator and
$\mu(t)=\int dx\ (\phi^{*}H\phi+g|\phi|^{4})$ is the time-dependent
chemical potential. The presence of the projection operator $Q=1-|\phi\rangle\langle\phi|$
ensures the total number-conservation of the particles~\cite{Castin:LowTemperatureBECsInTimeDependent:PRA98}. 
The condensate wave-function and the Bogoliubov 
mode amplitudes can be written as a Fourier 
series: $f(x) =  L^{-1}\sum_{l\in\mathbb{Z}} e^{ilx}\tilde{f}(l)$
and $\tilde{f}(l) = \int_0^L  e^{-ilx}f(x)dx$, 
where $f=\phi,u_k,v_k$. For an initial uniform distribution $\phi(x,t=0)=L^{-1/2}$,
the corresponding initial values of $(u_{k},v_{k})$, obtained from
the diagonalization of the operator $\mathcal{L}(t=0)$~
\footnote{The first kick being applied just after, at $t=0^{+}$.}, are plane waves of momentum $k$ (in units of $\kbar$)
\begin{equation}
\left[\begin{array}{c}
\tilde{u}_{k}(l,t=0)\\
\tilde{v}_{k}(l,t=0)
\end{array}\right]=\frac{\sqrt{L}}{2}\left[\begin{array}{c}
\xi+1/\xi\\
\xi-1/\xi
\end{array}\right]\delta_{k,l}\label{eq:uk0}
\end{equation}
with $k \in \mathbb{Z}^{*}$, and $\xi$ given by: 
\begin{equation}
\xi=\left[\frac{k^2}{k^2+2g/\pi\kbar^2}\right]^{1/4}.\label{eq:def_xi}
\end{equation}

We emphasize that the Bogoliubov modes $(u_k,v_k)$ are momentum eigenstates only at time $t=0$; once the kicks are applied, 
different components of the momentum distribution are mixed. 

Let us now consider
the stability of the condensate by estimating the number of non-condensed atoms  (the quantum depletion) at zero temperature
of the gas which is given by $\delta N = \sum_k N_k$, where the number of excitations $N_k$ in the mode $k$ is 
\begin{eqnarray}
N_{k}&=&\int_0^L|v_{k}(x)|{}^{2}dx=\frac{1}{L}\sum_l |\tilde{v}_{k}(l)|^{2}.\label{eq:Nk}
\end{eqnarray}
As the total number of particles is fixed, the number of condensed particles is $N-\delta N$ and the non-condensed fraction is simply $\delta N/N$. 
As long as $\delta N$ is much smaller that the typical number of atoms $\approx10^5$ used in a experiment, 
the kicked condensate is stable. 
Due to the inversion symmetry of the problem, we can restrict the study to $k>0$. We will focus on the 
initially most populated mode $k=1$.	

\begin{figure} 
\begin{centering}
\includegraphics[height=4.cm]{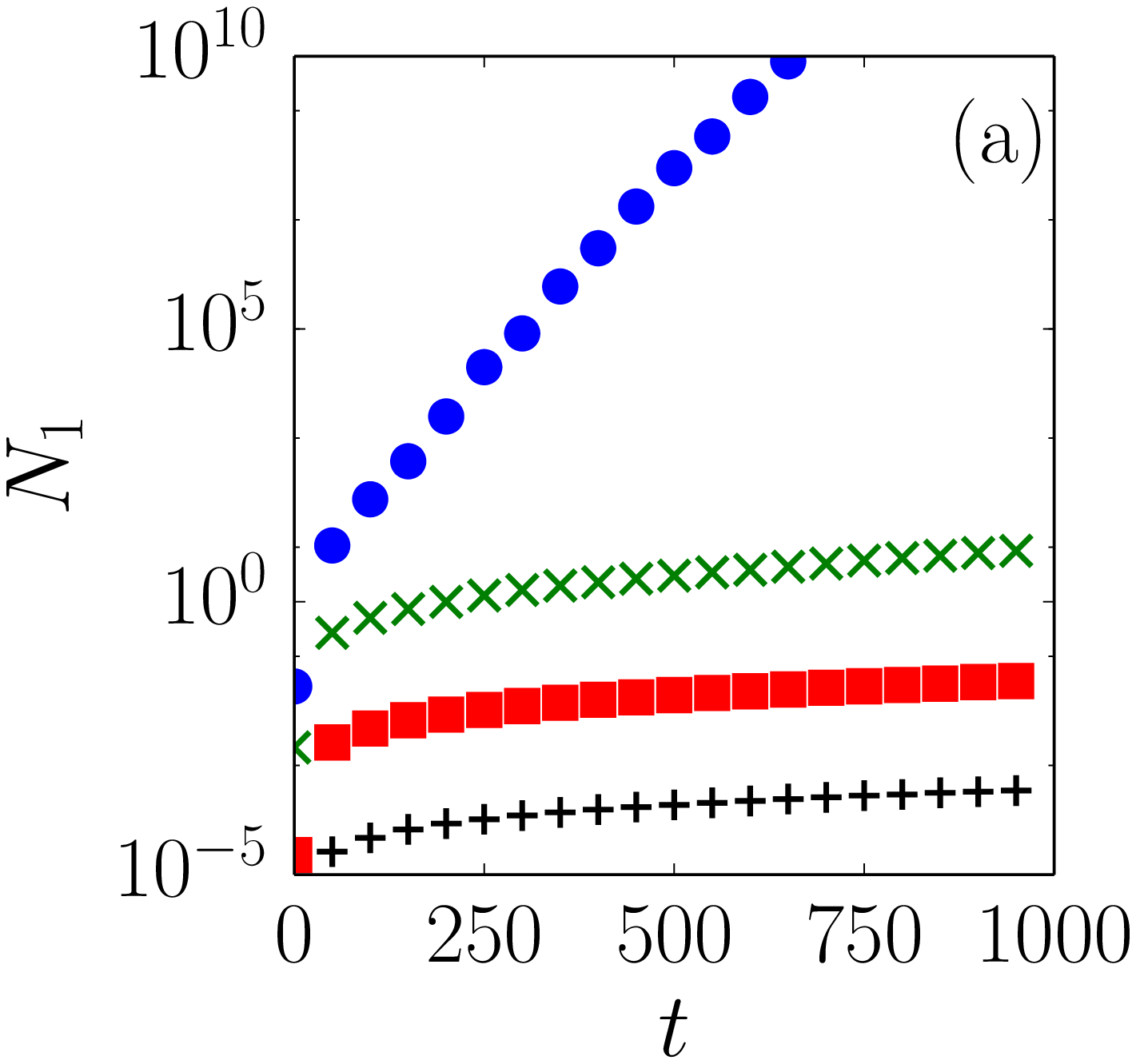} 
\includegraphics[height=4.cm]{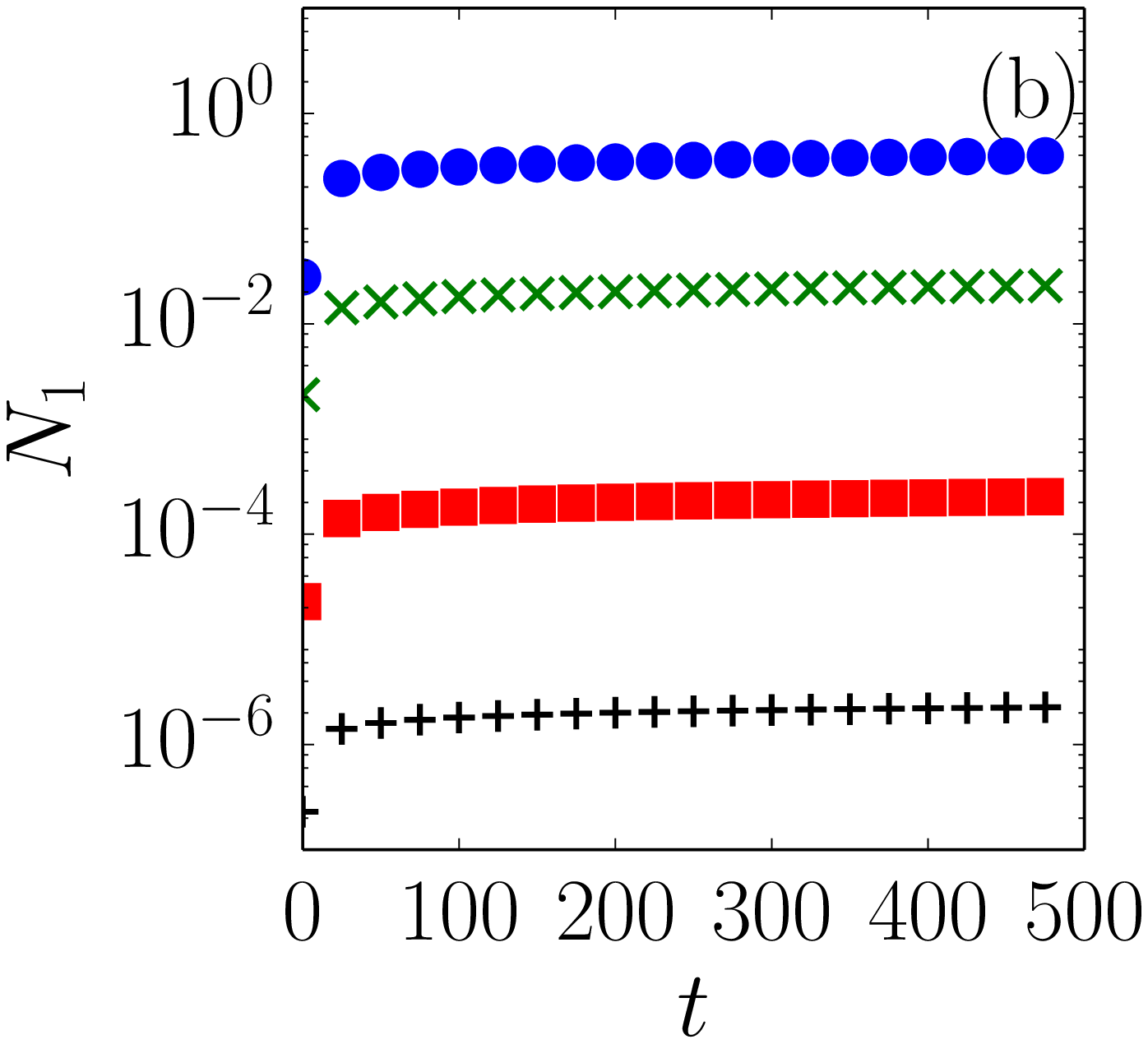}
\end{centering}

\caption{(Color online) Evolution of the population of the mode $k=1$ for
$g=10^{-2}$ (black $+$), $g=10^{-1}$ (red $\square$), $g=1$
(green $\times$) and $g=4$ (blue $\circ$) (a) in the quasilocalized regime
$K=4,\ \epsilon=0.1$ and (b) in the diffusive regime $K=9,\ \epsilon=0.8$
(b).
\label{fig:N1vst}}
\end{figure}

For the periodic kicked rotor, several studies showed the emergence of an instability at large positive values of $g$
(repulsive interactions) \cite{Zhang:TransitionToInstabilityKickedBEC:PRL04,*Billam:CoherenceAndInstabilityInADrivenBEC:NJP12,*Reslen:DynamicalInstabilityInKickedBEC:PRA08},
which manifests itself by an exponential increase of the number of
excitations. We shall
now study this instability in the {\it quasiperiodic} kicked rotor for $g>0$. Equations~\eqref{eq:evolphi}
and~\eqref{eq:evol_uk} can be integrated simultaneously by a split-step
method. Numerical data are averaged over 500 random realizations of the phases $\varphi_{2},\varphi_{3}\in[0,2\pi)$. Figure~\ref{fig:N1vst}(a)
displays the time evolution of $N_{1}$ for different values of $g$
in the \emph{quasilocalized} regime. For low interaction strengths
$g\le0.1$ the system is stable in the considered time range, the
average number of excitations $N_{1}\le0.1$ being very small.
For $g=1$ the interplay between the kicks and interactions leads
to a slow exponential increase of the number of quasiparticles:
$N_{1}(t=1000)\approx10$, and for $g=4$, the condensate is clearly unstable,
the number of excitations exceeding the typical number of condensed
atoms $N$ after a few hundred kicks. 
Figure~\ref{fig:N1vst}(b) shows the number of excitations for a larger kick amplitude $K=9$ with $\epsilon=0.8$ corresponding 
to the diffusive regime. In this case, the condensate is less affected by the presence of interactions, 
as the kinetic energy grows linearly with time and eventually dominates the
constant interaction energy $\simeq g$.
In the following, we thus focus on very low interacting
strengths $g\le0.1$ for which $N_{1}\ll N$ meaning that the condensate
is stable and the Bogoliubov formalism is valid.

\begin{figure*} 
\begin{centering}
\includegraphics[height=2.7cm]{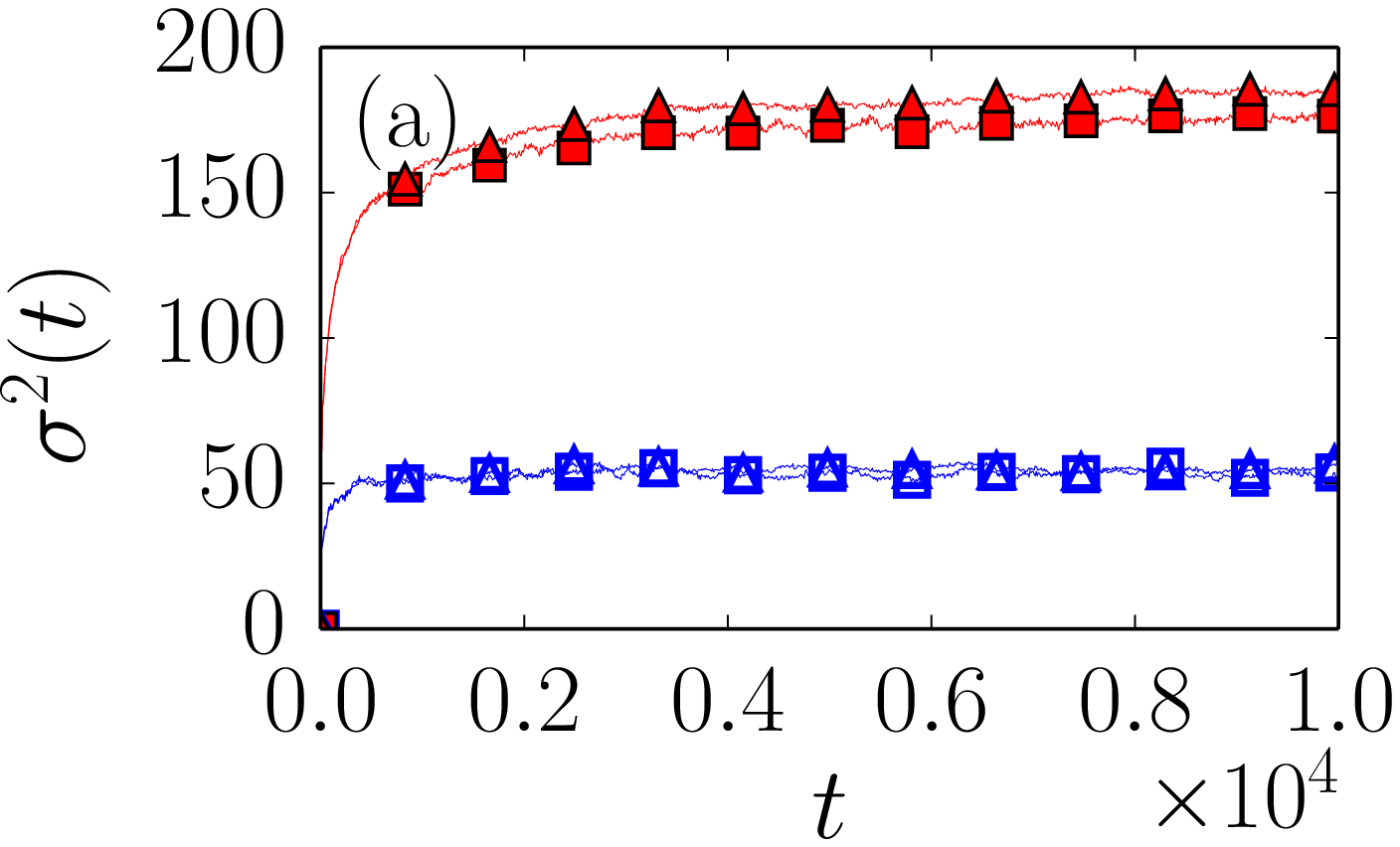} 
\includegraphics[height=2.7cm]{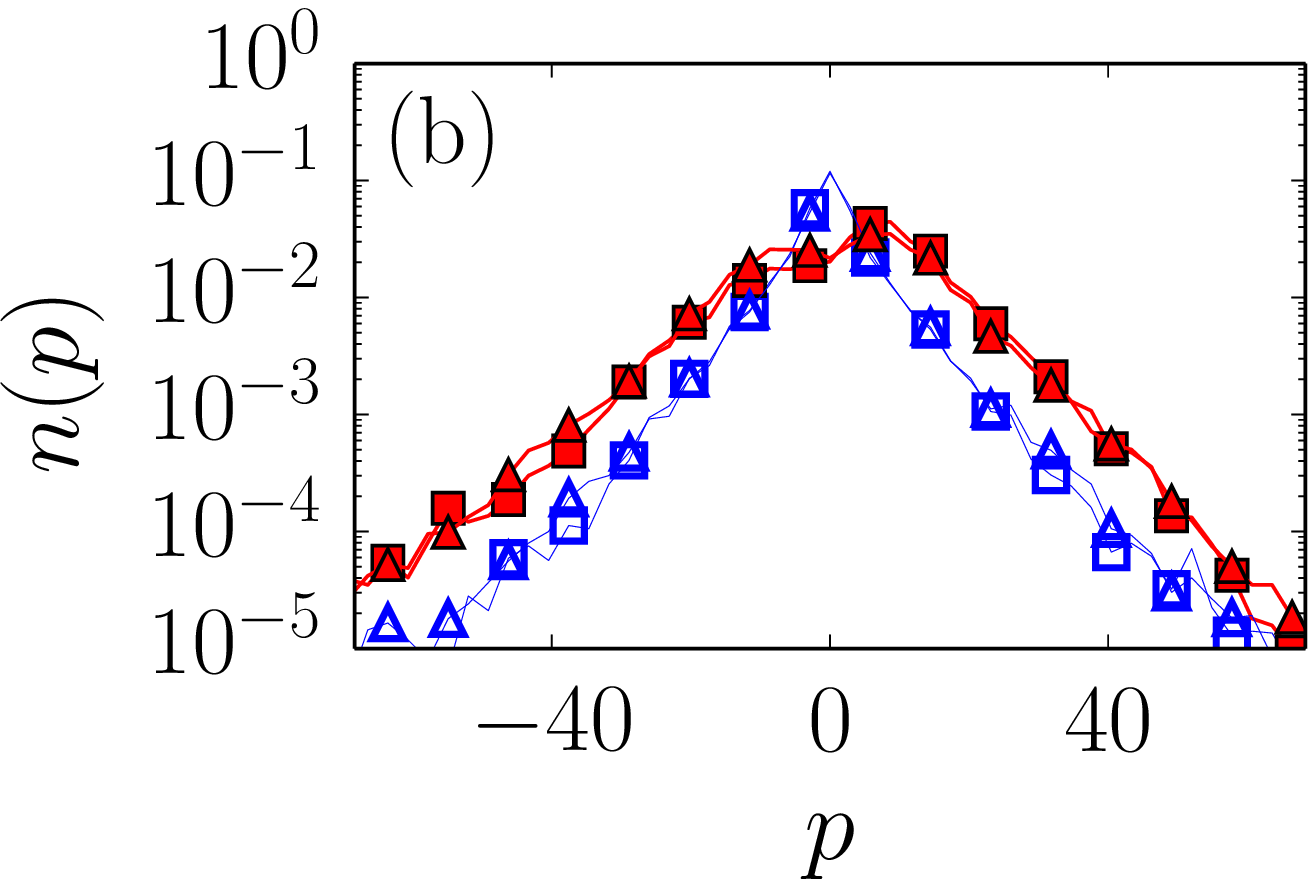}
\includegraphics[height=2.7cm]{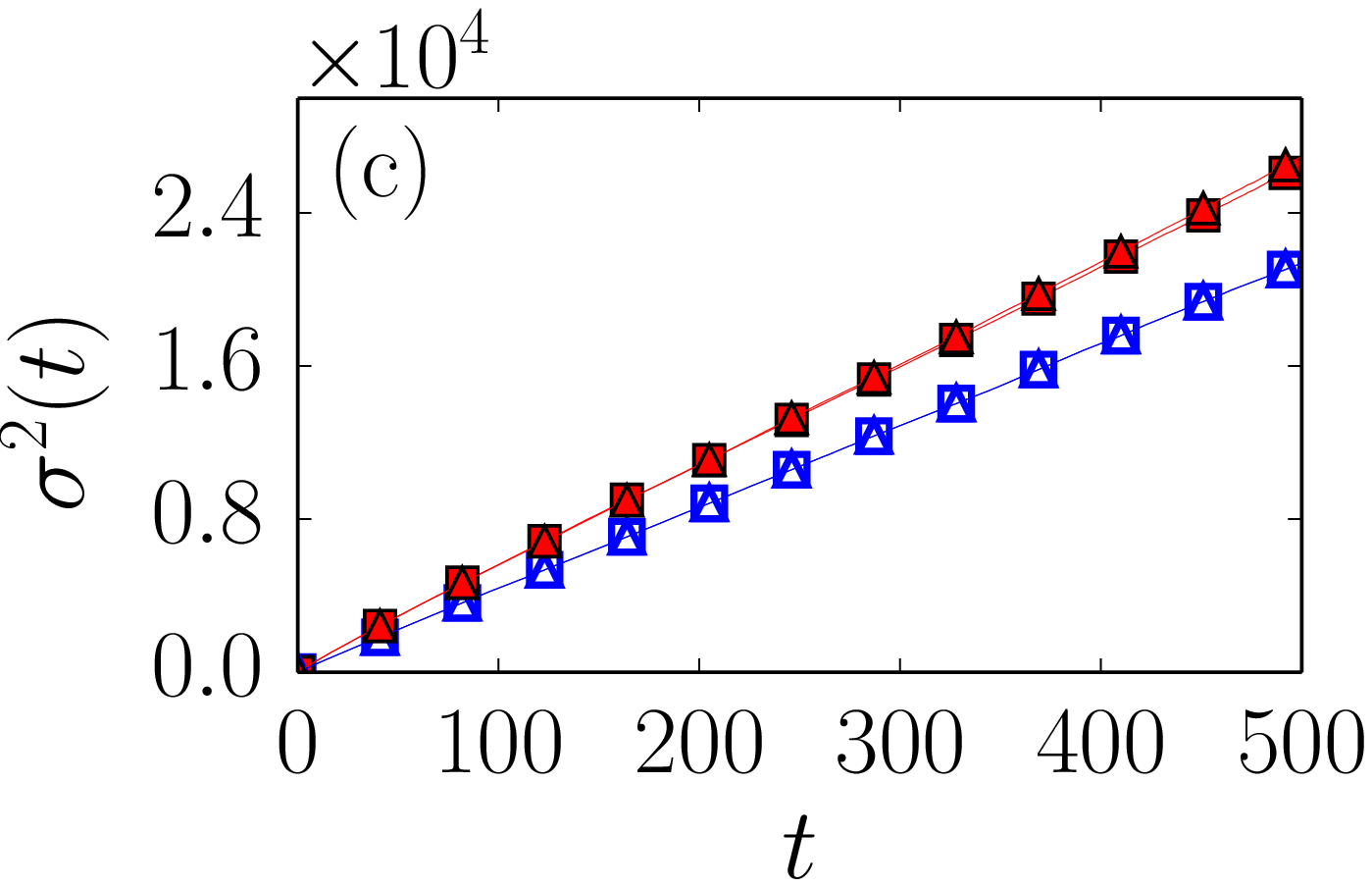} 
\includegraphics[height=2.7cm]{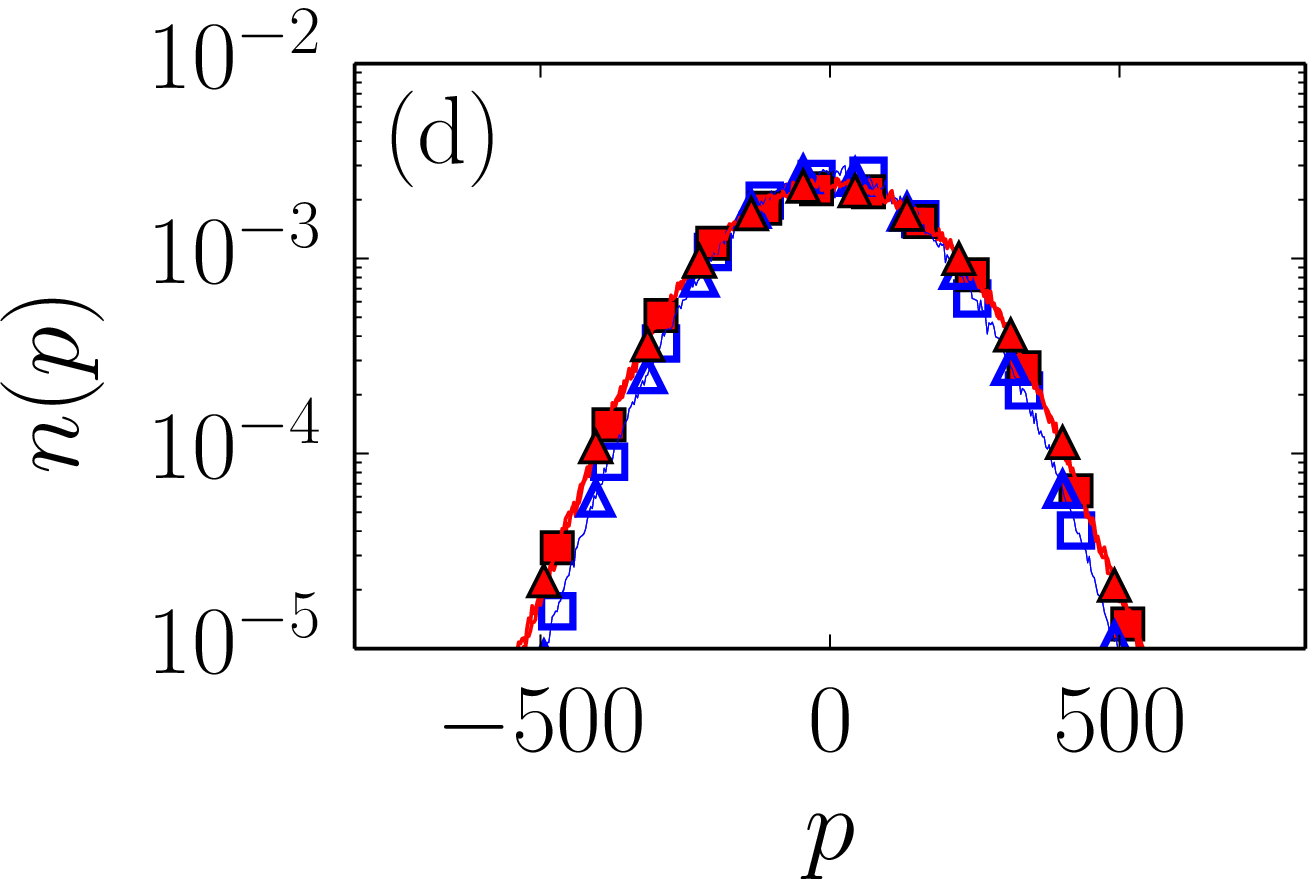}
\end{centering}

\caption{(Color online) Dynamics of the condensate and of the quasiparticles.
Variance of the momentum of the condensate $\sigma_{\mathrm{c}}^{2}(t)$
(empty blue markers) and of the quasiparticles $\sigma_{\mathrm{b}}^{2}(t)$
(full red markers) for $g=10^{-4}$ (triangles) and $g=10^{-1}$ (squares) in the
(a) quasilocalized regime ($K=4, \epsilon=0.1$) and (c) in the diffusive regime ($K=9, \epsilon=0.8$).
Momentum distributions of the condensate $n_{\mathrm{c}}(p)$ (empty markers) and of the quasiparticles
$n_{\mathrm{b}}(p)$ (full markers) at $t=10^{4}$ (logarithmic scale) in the
(b) quasilocalized regime and (d) diffusive regime.
\label{fig:profiles}}
\end{figure*}

We now consider the (normalized) momentum distribution of the condensate,
\mbox{$n_{\mathrm{c}}(p=\kbar l)=|\tilde{\phi}(l)|^{^{2}}/(L\kbar)$} 
and that of the quasiparticles, \mbox{$n_\mathrm{b}(p=\kbar l)=\left|\tilde{v}_{1}(l)\right|^{2}/(L N_1 \kbar)$}.
For $g\le0.1$ and for short enough times, one expects the
condensate to display quasilocalization if $K<\tilde{K}$ and diffusion if $K>\tilde{K}$,
$\tilde{K}$ being the critical point. 
Our numerical simulations show that
this is also the case for the quasiparticles. We have checked that momentum distributions remain 
essentially centered around the origin so that their first moments $\langle p \rangle_i=\kbar^2 \sum_l l n_i(p)$ ($i=\mathrm{c,b}$) are small. 
Figure~\ref{fig:profiles}(a) shows
the second moment of the distribution variance of the distribution $\sigma^2_{i}=\langle p^2\rangle_i-\langle p \rangle_i^2$, with $\langle p^2\rangle_i=\kbar^3 \sum l^2 n_{i}(p)$, 
for both the condensate ($i=\mathrm{c}$, blue empty markers) and of the excitations
($i=\mathrm{b}$, red full markers) in the quasilocalized regime. For the two values
of the interacting strength $g=10^{-4}$ (triangles) and $g=10^{-1}$
(squares), the second moment of the condensate saturates to a constant
value $\sigma_{\mathrm{c}}^{2}=50$, showing that the wave-packet is quasilocalized. 
Assuming an exponential profile $n_\mathrm{c}(p)\propto\exp(-|p|/\xi)$,
the localization length~\footnote{As commented above, in presence of interactions there is, strictly speaking, 
no real localization, but we shall call $\xi$ ``localization length'' for simplicity.}
 of the momentum distribution at $t=10^4$ is given by $\xi=\sigma_{\mathrm{c}}/\sqrt{2}\sim 5$,
which evolves very slowly with time up to $t=10^4$ [cf. Fig.~\ref{fig:profiles}(a)]. In other
words, for very weak interactions and at the short time scales accessible to experiments, the condensate behaves as a single particle
and displays a behavior very close to the Anderson localization. More interestingly, Fig.~\ref{fig:profiles}(a)
shows that \emph{Bogoliubov quasiparticles also tend to ``localize"},
although with a larger value of the second moment $\sigma_{\mathrm{b}}^{2}\approx180$.
The momentum distributions in the quasilocalized regime at $t=10^{4}$ are shown in Fig.~\ref{fig:profiles}(b)
which displays $n_{\mathrm{c}}$ and $n_{\mathrm{b}}$ [same
graphical conventions as in Fig.~\ref{fig:profiles}(a)]. The momentum distribution of the condensate has a typical exponential
profile associated with Anderson localization, whereas that of the quasiparticles presents two peaks with
exponential wings. This peculiar shape is probably due to the asymmetric
initial condition, the initial momentum distribution of the mode $k=1$
[Eq.~\eqref{eq:uk0}] is centered at $p=\kbar$.
The wings of the quasiparticle momentum distribution have approximately the
same slope as the condensate distribution, showing that quasiparticles have the same localization length 
in this example. Because $n_\mathrm{b}(p)$ has a much flatter top than 
$n_\mathrm{c}(p),$ the associated second moment $\sigma_{\mathrm{b}}$ is however significantly larger than $\sigma_{\mathrm{c}}.$

In the diffusive regime, the condensate and the quasiparticles have the same dynamical behavior: Figure~\ref{fig:profiles}(c), which
is the equivalent of Fig.~\ref{fig:profiles}(a) for $K=9,\ \epsilon=0.8$,
actually shows that $\sigma_{\mathrm{c}}^{2}$ and $\sigma_{\mathrm{b}}^{2}$
increase linearly with time and that the evolution is very similar for $g=10^{-4}$ and $g=10^{-1}.$ The diffusion coefficients $D_\mathrm{c}=\sigma_{\mathrm{c}}^{2}/(2t)\sim20$
and $D_\mathrm{b}=\sigma_\mathrm{b}^{2}/(2t)\sim25$ are also similar. Figure~\ref{fig:profiles}(d)
represents the corresponding momentum distributions at $t=500$. Both have the typical Gaussian shape associated with a diffusion process.

These results show that the nature of the two phases of the Anderson
metal-insulator transition in the QPKR are not changed, at least for
relatively short times, by weak interactions.

\begin{figure} 
\begin{centering}
\includegraphics[height=4cm]{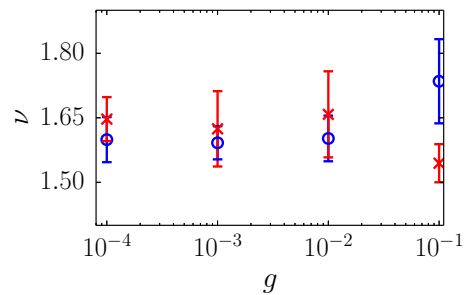}
\end{centering}

\caption{(Color online) Critical
exponent $\nu$ {\it vs} interacting strength $g$ for both the condensed fraction
(blue circles) and the Bogoliubov quasiparticles (red crosses). Error bars are calculated via a standard bootstrap method~\cite{NumRec:07}.
\label{fig:critical}}
\end{figure}

We now study the critical behavior. As Bogoliubov quasiparticles 
behave like real (non-interacting) particles in the quasilocalized
and diffusive regimes, it is reasonable to expect that they also
display the Anderson transition. The universality of this second-order
phase-transition has been recently demonstrated experimentally in
the absence of interactions~\cite{Lopez:ExperimentalTestOfUniversality:PRL12},
by showing that the experimental value of the critical exponent $\nu=1.63\pm0.05$
is independent of the microscopic parameters of the system, and consistent
with the numerically predicted value $1.58\pm0.02$~\cite{Lemarie:UnivAnderson:EPL09,*Slevin:AndersonCriticalExp:NJP2014}.
In order to access the critical properties of the system, we use a finite-time scaling method~\cite{Lemarie:AndersonLong:PRA09,*Slevin:AndersonCriticalExp:NJP2014}
which allows one to characterize the critical regime of the phase transition
and to extract its critical exponent $\nu$. The critical regime is shown to correspond to a subdiffusive
expansion with exponent $2/3$: $\sigma^{2}\propto t^{2/3}$~\cite{Lemarie:AndersonLong:PRA09}. 
We cross the transition along the path $\epsilon(K)=0.1+0.14(K-4)$ used
in~\cite{Chabe:Anderson:PRL08}. 
Figure~\ref{fig:critical} shows that, for small nonlinearities, the critical exponent is the same
for both components and compares very well with the (non-interacting)
experimental measurement $\nu=1.63\pm0.05$~\cite{Lopez:ExperimentalTestOfUniversality:PRL12},
but their values tend to become different for higher values of $g$. The critical point is also found to be the
same for both the condensate and the excitations. Its value, $\tilde{K}\approx 6.38\pm0.05$
at $g=0$, practically does not change (inside the statistical uncertainty) up to $g=0.1$, in accordance with the self-consistent theory prediction of
\cite{Cherroret:AndersonNonlinearInteractions:PRL14}.
This is the main result of the present work: Bogoliubov quasiparticles undergo a second-order phase transition and 
the corresponding critical exponent has the same value as the one observed
for a non-interacting system of independent particles. This suggests that the concept of universality is valid for independent
particles, for interacting condensates and for Bogoliubov quasiparticles: All
these conceptually different objects undergo a phase transition with
the \emph{same} critical exponent. For $g \ge 0.1$, the value of the critical exponents 
starts to deviate from the universal value illustrating that the system enters a new regime where 
the kicked condensate is changed by the presence of interactions,
that is, the subdiffusive character of the quasilocalized regime becomes important even for short 
times~\cite{Cherroret:AndersonNonlinearInteractions:PRL14}.

The above study is restricted to the Bogoliubov mode $k=1$. 
Experimentally, Bogoliubov modes can be selectively excited
using two laser waves whose directions are chosen so that their
wave-vector difference  $\Delta k_L$ corresponds to the wave-vector $k$ of the desired
mode~\cite{Steinhauer:ExcitationSpectrumBEC:PRL02}.
Considering another mode $k\neq1$ is equivalent to a change of the initial
condition in the Bogoliubov equations and should not change our results.
We checked numerically that theses modes, which are initially less
populated than the $k=1$ mode, display the same behavior, but are
much more affected by finite-time effects, as their initial momentum
distribution is more asymmetric [see Eq.~\eqref{eq:uk0}].  
We also mention that it would be 
interesting to extend the present study to higher values of $g$ and/or longer time scales where dynamical localization is supposed to be effectively destroyed, and see if quasiparticles display
the same subdiffusive dynamics {\it below} the critical point, as predicted for the condensate
by the self-consistent theory \cite{Cherroret:AndersonNonlinearInteractions:PRL14},
but such a study is outside the scope of the present work.

In conclusion, in a quasiperiodic kicked rotor, in the very weakly
interacting regime, the condensate is stable for times that are larger than 
the experimental time-scale (presently up to 1000 kicks). 
In this regime, both quasiparticles and the condensate
behave like single particles undergoing the Anderson transition from
a localized regime to a diffusive regime, and display the same critical
exponent, which is compatible with the one observed for non-interacting particles; the universality of the
phase transition is thus valid irrespectively of the type of particle.
The above findings confirm the potentialities of the quasiperiodic kicked rotor for
the experimental study of the effect of interactions on the Anderson transition. For low positive values of $g$
the noninteracting regime can be experimentally observed, and, by increasing interactions
via a Feshbach resonance one can observe the onset of nonlinear effects. 
The present work opens the way for such an experiment, which would represent an important
advance in the physics of interacting disordered systems.

\begin{acknowledgments}
The authors are grateful to  R.~Chicireanu, J.-F. Cl\'{e}ment, P.~Szriftgiser,
and N.~Cherroret for fruitful discussions. JCG thanks the
Max-Planck Institute for the Physics of Complex Systems, Dresden,
for support in the framework of the Advanced Study Group on Optical Rare Events.
Work partially
supported by Agence Nationale de la Recherche (grants LAKRIDI ANR-11-BS04-0003 and K-BEC ANR-13-
BS04-0001) and the Labex CEMPI (ANR-11-LABX-0007-01).
\end{acknowledgments}

\bibliographystyle{apsrev4-1}
\bibliography{ArtDataBase}

\end{document}